# Artificial Intelligence-Assisted Workflow for Transmission Electron Microscopy: From Data Analysis Automation to Materials Knowledge Unveiling

Marc Botifoll[1,*,†], Ivan Pinto-Huguet[1,†], Enzo Rotunno[2,*], Thomas Galvani[1], Catalina Coll[1], Payam Habibzadeh Kavkani[2,3], Maria Chiara Spadaro[4,5], Yann-Michel Niquet[6], Martin Børstad Eriksen[7], Sara Mart´ı-Sánchez[1], Georgios Katsaros[8], Giordano Scappucci[9], Peter Krogstrup[10], Giovanni Isella[11], Andreu Cabot[12,13], Gonzalo Merino[7], Pablo Ordejón[1], Stephan Roche[1,13], Vincenzo Grillo[2], and Jordi Arbiol[1,13,*]

[1]Catalan Institute of Nanoscience and Nanotechnology (ICN2), CSIC and BIST, Campus UAB, Bellaterra, 08193, Barcelona, Catalonia, Spain
[2]CNR Istituto Nanoscienze, Via Campi 213/A, 41125 Modena, Italy
[3]University of Modena and Reggio Emilia, 42122, Reggio Emilia RE, Italia
[4]Department of Physics and Astronomy "Ettore Majorana", University of Catania, via S. Sofia 64, 95123 Catania, Italy
[5]CNR-IMM, via S. Sofia 64, 95123 Catania, Italy
[6]Univ. Grenoble Alpes, CEA, IRIG-MEM-L Sim, Grenoble, France
[7]PIC, IFAE, Campus UAB, Bellaterra, 08193, Barcelona, Catalonia, Spain
[8]Institute of Science and Technology Austria, Klosterneuburg, Austria
[9]QuTech and Kavli Institute of Nanoscience, Delft University of Technology, PO Box 5046, 2600 GA, Delft, The Netherlands
[10]NNF Quantum Computing Programme, Niels Bohr Institute, University of Copenhagen, Denmark
[11]Laboratory for Epitaxial Nanostructures on Silicon and Spintronics, Physics Department, Politecnico di Milano, Como, Italy
[12]Catalonia Institute for Energy Research – IREC, Sant Adria` de Beso`s, Barcelona, 08930 Spain
[13]ICREA, Passeig Llu´ıs Companys 23, 08010, Barcelona, Catalonia, Spain

*Corresponding authors: marc.botifoll@icn2.cat, enzo.rotunno@nano.cnr.it, arbiol@icrea.cat; † Equal contribution

**Abstract**

(Scanning) transmission electron microscopy ((S)TEM) has significantly advanced materials science but faces challenges in correlating precise atomic structure information with the functional properties of devices due to its time-intensive nature. To address this, we introduce an analytical workflow for the holistic characterization, modelling, and simulation of device heterostructures. This workflow automates the experimental (S)TEM data analysis, providing an in-depth characterization of crystallographic information, 3D orientation, elemental composition, and strain distribution. It reduces a process that typically takes days for a trained human into an automatic routine solved in minutes. Utilizing a physics-guided artificial intelligence model, it generates representative descriptions of materials and samples. The workflow culminates in creating digital twins—3D finite element and atomic models of millions of atoms—enabling simulations that provide crucial insights into device behaviour in practical applications. Demonstrated with SiGe planar heterostructures for scalable spin qubits, the workflow links digital twins to theoretical properties, revealing how atomic structure impacts materials and functional properties such as spatially-resolved phononic or electronic characteristics, or (inverse) spin orbit lengths. The versatility of our workflow is demonstrated through its application to a wide array of materials systems, device configurations, and sample morphologies.

# 1 Introduction

In an era marked by profound digital transformations, semiconductor heterostructures within a chip have emerged as crucial and widespread assets, driving major industrial value chains. They support advancements in both novel sectors (such as automated vehicles, cloud computing, Internet of Things, space exploration, supercomputing, and quantum technologies) and traditional ones (including computing and communications, industrial automation, entertainment, and healthcare). [1, 2] Miniaturisation, now reaching the nanoscale and approaching the atomic limit, stands out as a primary driver of progress. This trend enhances device capabilities, lowers costs, and reduces energy consumption. [3]

The tackling of these technologies and their required extreme miniaturisation signifies a paradigm shift in device design, where “every atom matters”. In contrast to traditional electronic devices, the diminishing sizes introduce significant nuances, such as fluctuations in dopant concentration, interdiffusion at interfaces, and local strain fields at the nanoscale, profoundly impacting device function and performance. [4, 5] Devices for quantum computing exemplify the extreme case. Structural deviations from the ideal conceptual design may have noticeable effects in the quantum performance, even though the exact causes and correlations between these structural features and the final functional properties are still unknown in most systems. [6] This ubiquitously occurs at different scales in material systems in which miniaturisation is key towards property improvement, and precise characterisation of every single atom will thus be key in understanding its properties when embedded in a full device (i.e., ferroelectrics and their interatomic distances, catalytic particles and their progression into single-atom catalysts, batteries and the atom-wise *in-situ* analysis of their degradation mechanisms, and others). [7, 8, 9, 10, 11, 12]

Beyond their characterisation, the exploration and development of novel materials and devices, as well as the optimisation of existing ones for various applications, constitute a multifaceted process involving the identification of needs to fulfill, literature review, material proposal, device engineering, characterisation, and application testing. [13] This iterative cycle, driven by multidisciplinarity and collaborative efforts, forms the backbone of scientific progress. However, it is often hindered by its time-consuming and expensive nature, particularly when atomic scale precision is required for understanding and controlling the functionality of materials and heterostructured devices. The gold standard for achieving this atomic characterisation is (Scanning) Transmission Electron Microscopy ((S)TEM). However, this technique is currently constrained by the aforementioned challenges, which restrict the number of experimental repetitions and diminish the statistical significance of the results, which is the backbone for technological progress in the microelectronic industry. [14, 15, 16, 17]

While significant strides have been made in automating (S)TEM data acquisition, especially in industrial settings for metrology and process characterisation, a substantial challenge persists in extracting meaningful physical insights from the vast amount of data (raw images and spectra) generated during experimental analysis. [18, 19] This challenge hinders a deep understanding of embedded material heterostructures in devices at the atomic level. Conventional fab and lab metrology tools fall short in providing a comprehensive and efficient analysis of these intricate device architectures, leading to a lack of statistical sampling for understanding performance variability among individual devices. As a result, the demand for high-throughput data analyses that provide statistical significance and link structural characterisation with functional properties is more justified than ever.

Therefore, it is fundamental to address the inherent challenges in the traditional (S)TEM-based materials exploration process, which can be facilitated by recent breakthroughs in data analysis. [20, 21, 22] Machine learning (ML), deep learning (DL), computer vision (CV), and artificial intelligence (AI), have transformed nearly every facet of our daily lives, and materials science is not an exception, enabling levels of accuracy, precision, and noise tolerance previously considered unachievable in (S)TEM-related analyses. [17, 18, 23, 24] However, since their introduction in electron microscopy for materials science, the challenge of generalising their methods has been regarded as its greatest limitation.

The early stages of AI-driven methodologies associated with (S)TEM data analysis, characterised by the utilisation of unsupervised unmixing algorithms to decompose hyperspectral signals like Energy Dispersive X-Ray Spectroscopy (EDX) or Electron Energy Loss Spectroscopy (EELS) spectra, have evolved into the adoption of advanced models such as convolutional neural networks, autoencoders, or reinforcement learning. [17, 25, 26, 27, 28, 29] These advanced models are employed to unveil and learn features from images and high-dimensionality signals such as spectral images or 4D-STEM data. [30, 31] Notably, these advancements have paved the way towards the automation of both experiments and data analysis.[32, 33, 34, 35, 36] These innovations result in unparalleled insights from AI-based data analysis while still constrained by case-specific routines and limited statistical significance. [37, 38, 39, 40, 41]

In the present manuscript, we explore the integration of these advanced techniques into a comprehensive and automated characterisation workflow, aiming to overcome the traditionally slow and tedious aspects of materials characterisation. Not only does our solution provide a new paradigm of automation in STEM data analysis but also an easy way to generate new knowledge from representative 3D models of the experimental devices which would otherwise imply an unpractical amount of manual inputs (atom-by-atom in atomic models, and contour-by-contour in finite element models). The proposed workflow starts by automating the data analysis process, traditionally considered a bottleneck, turning it into a solution for rapid and reproducible knowledge retrieval. The manuscript outlines a step-by-step approach inspired by the logical progression of human microscopists, beginning with low-magnification segmentation to gauge device morphology and culminating in phase and orientation-sensitive Geometrical Phase Analysis (GPA), for detailed structural insights on local atomic displacements, strain and defects. [42, 43] The workflow extends beyond data analysis and incorporates the automated generation of representative 3D atomic (3DAMs) and Finite Element (FEMs) models, utilizing the experimental data collected. Notably, these models comprehensively capture all the experimental information obtained through the preceding automated steps. As a result, the automated workflow can simulate a device that closely matches the originally designed, grown, and engineered device (i.e., a digital twin). We refer to this process as an “experimental simulation”, as the models are created from experimental data and parameters. For example, the finite elements of the FEMs are derived from the contours outlined by the segmentation of low-magnification images, while the atomic positions in the 3DAMs can be determined by the displacements identified through GPA on atomic resolution STEM images. This capability facilitates an efficient workflow and enables the exploration of atomistic models (digital twins) comprising millions of atoms. The strain from these models have been studied by finite element relaxations and atomic Keating models, which can also address the vibrational properties of the material, while their electronic structures are derived from precise 3DAM-based tight-binding Hamiltonians and computed using linear scaling algorithms. [44, 45] Ultimately, this establishes a direct link between realistic atomistic representations of as-grown materials and the variations in both local and global physical properties of the associated quantum devices.

The proposed workflow is flexible and adaptable due to its modular nature, whose interconnection transcends the sum of its parts, which also individually push the state of the art. This modularity allows seamless integration of cutting-edge research and open-source tools from diverse sources. The potential applications of this automated approach extend to diverse fields, including energy and environmental research, classical communications, quantum technologies, mechanical engineering, and fundamental chemical research.

In subsequent sections, we delve into the details of each step within the proposed workflow, elucidating the interconnectedness and modularity that make this method a promising avenue for accelerating materials science research. The manuscript also discusses the broader implications of this workflow and its potential to reshape the landscape of materials exploration and characterisation. [18, 46]

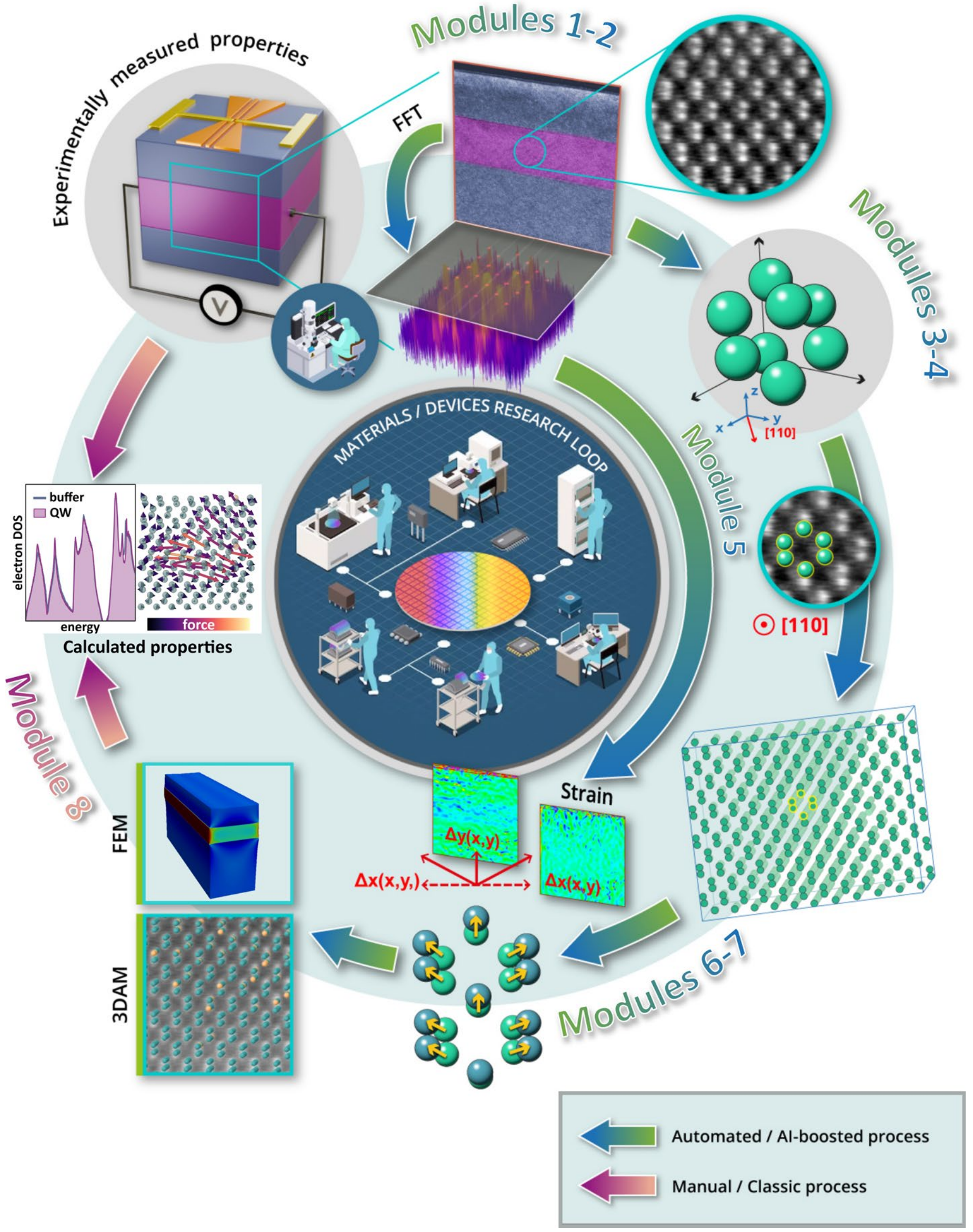

Figure 1: Scheme of the proposed workflow: Module 1: Initially, images are segmented, breaking them down into distinct regions. Modules 2-4: The micrograph analysis in Fourier space characterises the crystallography within each segmented region. Module 5: This comprehensive local crystallographic characterisation enables to map strain and to get finer structural details, such as interfaces or defects. Modules 6-7: The compiled information is then transferred to create Finite Element and 3D Atomic Models (FEM/3DAM). Mpdule 8: These models, in turn, serve as the basis for computing functional properties like the electronic and vibrational states, or atomic forces and strain relaxations. This sequential process ensures a comprehensive analysis and understanding of the material's functional characteristics and behaviour.

# 2 Workflow

The workflow connects the data that is obtained directly from the electron microscope with the properties of the imaged heterostructure or device, in an intuitive and automated manner. This paper concentrates on automating the analysis based on single images or stacks of them, although compositional information is incorporated when necessary to enhance accuracy.

The choice between a single micrograph and a stack of images depends on the device's or heterostructure's size, imaged features, and microscope sampling. A single micrograph is suitable when it captures both the morphology and the atomically-resolved structural information. Conversely, a stack of images is used for larger devices or structures, requiring a progressive increase in magnification to transition from the morphological overview to atomic and structural details. Both situations are handled differently, with the latter requiring an additional processing step to link information gathered at each magnification.

The modular workflow is represented in figure 1. The full cycle essentially consists of eight main modules which sequentially 1) segment the micrographs; 2) peak-find the Fast Fourier Transform (FFT); 3) identify the crystal phase from a single image or 4) a stack of them; 5) compute strain; 6) build representative 3D models of the experimental sample or device, either FEMs or 7) 3DAMs; and finally 8) calculate functional properties based on the model representation of the experimental device. These modules are self-sufficient and can independently produce results on their own, but also be autonomously interconnected to link their outputs into a final result of additive complexity.

Every independent module, its contribution to the state of the art, and its additive complexity towards our proposed "experimental simulations" are described next.

## 2.1 Segmentation

Initially, our goal is to reveal the morphology of heterostructures and devices, encompassing their size, the identification of key regions, their spatial distribution and their interfaces and contours. These regions may have different chemical compositions, crystallographic phases or orientations. Image contrast (differences in pixel intensity) raised by the electron microscope and its post-processing allows us to separate these regions as segmented units, which is particularly challenging given the study's aim for a general solution applicable to a broad range of devices and samples. [47, 48] This targeted diversity and the consequent absence of labelled data has required an unsupervised machine learning approach. [49, 50, 51, 52] As detailed in the supplementary information, section 1.1 entitled "Segmentation", among the available state-of-the-art segmentation methods, we chose the Canny edge detection algorithm, which we optimized, automated, and generalised for our target data type, [52], as well as the state-of-the-art general-purpose segmentation neural network Segment Anything Model (SAM). [53] Having both integrated into the workflow overcomes individual limitations such as segmentation scenarios where feature edges are represented by blurry intensity gradients, such as parallel beam TEM micrographs, where Canny edge detection would struggle (Fig. S3 and S4). [53] SAM allowed us to expand the domains of our targeted systems to parallel beam TEM data whose diffraction contrast hinders an intuitive naked-eye segmentation. [53, 54] The optimised Canny edge detection model is tailored for heterostructures to detect (S)TEM edges (in high angle annular dark field (HAADF), bright field (BF) or integrated differential phase contrast (iDPC) STEM, BF TEM and high-resolution TEM (HRTEM) imaging modes) comprising dozens of pixels while SAM's general digital image processing focuses on intensity gradients happening in less than five pixels. Regarding computing efficiency, Canny edge model stands as a faster solution compared to SAM. Specifically, up to two orders of magnitude faster in demanding samples where SAM categorises the edges as additional segments and demands an additional processing step to merge these contours with the actual segments. The workflow offers the selection of the segmentation model based on the targeted device, highlighting the benefits of the human-in-the-loop approach to add control layers to a fully automated process, if needed. Moreover, the modularity of the workflow easily enables to incorporate alternative segmentation models from the literature to tailor it to each case, e.g. optimising it for nanoparticle analysis. [18] Comprehensive details of the model performance, training, metrics, labelling processes, and more, are available in the supplementary information, section 1.1.3 entitled "Final segmentation model proposed".

## 2.2 Fast Fourier Transform peak finding

After segmentation, the crystallographic information, encoded in reciprocal space, must be automatically extracted from the Fast Fourier Transform (FFT). [55, 56, 57] The key is using the segmentation to spatially filter the information represented in the FFT, either by masking or by cropping from the segmentation (supplementary information, section 1.2 "Peak finding in the Fourier spectra"). The goal in either case is to find the reciprocal space coordinates of the frequency peaks in the FFT representing the periodicity of crystallographic planes, as they encode the information of the imaged local crystallographic phases. The workflow can differentiate between amorphous regions, single crystals, and polycrystalline materials (Fig. S8, S9 and S10). Therefore, we designed our peak-finding method to be robust enough to maximise the detection of frequency spots in the reciprocal space (FFT) corresponding to crystallographic planes (recall: % of detected real planes from all planes appearing in the image) without noise (precision: % of real planes from everything detected as planes) (Fig. S12 and S13). Three or more detected planes per crystallographic phase are enough to instill confidence in model-based phase identification. Therefore, we prioritise maximising recall over precision to minimise the possibility of missing local crystallographic phases represented by a reduced number of planes in the FFT.

The existing peak-finding algorithms are abundant. [58, 59, 60, 61] Nevertheless, the available methods so far imply manual hyperparameter fine-tuning and sample-dependent considerations, for which we propose a fully automated and structure-independent solution. Our peak finding model relies on the successive application and evaluation of up to three distinct methods: 1) experimental ML-based 2D Gaussian fitting, 2) a trained U-Net model on synthetic data, and 3) a 1D profile scanning. [47, 59] The first and second methods lie within supervised ML/DL, while the third can be regarded as a computer vision algorithm. First, the ML-based 2D Gaussian fitting automatically optimises the parameters that would classically demand manual tuning in a 2D Gaussian fitting. [62] Second, a custom model based on noisy kinematical diffraction patterns is deployed to train a U-Net model peak (spots as crystal planes) identifier. [34, 47] The third and complementary 1D profile scanning method parallelly scans the FFT vertically and horizontally, merging both outputs to eliminate misleading cross-shape artefacts. These three approaches avoided manual labelling approaches to rely on model-based labelling.

The individual performance of each of the three methods could not satisfy the recall-precision balance required to successfully perform the upcoming modules of the workflow in the wide range of tested materials systems, heterostructures and devices. Their performance is detailed in the supplementary information, section 1.2.5 "Peak finding performance metrics". Therefore, we combined the three through a pipeline capable of, first, detecting if the material is amorphous or crystalline, and second, maximising the recall to deal with crystalline samples with multiple identifiable spots. While not flawless, it achieves the optimal balance between well-identified peaks and false positives and demonstrates adaptability across a wide range of materials and corresponding image types (Figs. S8-10, S12 and S13).

By testing this global model with 1000 manually labelled experimental FFTs from multiple materials, geometries and crystalline configurations, its precision of 69.78%, recall of 70.89%, and F1 score of 61.87%, make it the most robust automated peak finding model available (details on metrics available in the supplementary information, section 1.2.5 "Peak finding performance metrics"). The recall and precision are high enough to ensure that the lowest order Laue zones are well-identified, entailing a successful crystal phase identification, which is the eventual goal of the peak finding. Moreover, it detects neighbouring spots typically accounting for mismatched heterostructures or defects, which will be accounted for in the following steps. The nature and details of the peak-finding algorithm are carefully presented in the supplementary information, section 1.2.4 "Global combined peak finding model". However, the refinement, further implications and added potential of the proposed peak finding model are out of the scope of the present text and will be discussed elsewhere.

## 2.3 Phase identification

Once the crystallographic planes are located in the FFT, we can evaluate their goodness of fit with a database of candidate unit cells to match a crystallographic phase and its 3D orientation. To do so, a physics-aware model is convenient. Thus, this process involves ranking each potential crystal phase and assigning a score based on how well the experimental diffraction pattern encoded in the FFT aligns with their theoretical kinematical diffraction. [63, 64] The candidate unit cells are extracted from a crystallography database and filtered by prior knowledge (i.e., involved chemical species) to optimise the phase identification. [65, 66, 67, 68] The key lies in the iterative and combinatorial comparison of every pair of crystal planes detected in the experimental FFT with the candidate phases' theoretical pair of diffracted planes. By treating the problem with spot pairs as the validating unit, we make the phase identification robust and sensitive to structural defects that introduce additional planes in the FFT like twin boundaries or stacking faults. The output is an automated plane-wise indexation of the FFT and the most likely crystallographic phases considered from the database and their 3D orientation. The workflow incorporates an interactive graphical interface for visualising the raw FFT indexation and the sorted list of likely phases. The nature of the user interfaces can be found in section 2 "Proofs of the automated phase identification" of the supplementary information, as well as in the supplementary audiovisual material (videos showing the workflow in real-time).

The sequential segmentation and peak-finding enable the distinctive identification of peaks that would be too close together in a single global FFT of combined regions, allowing for separate frequency and phase identification. Consequently, the algorithm is resistant to cumulative spot position shifts caused by drift, or sensitive to subtle lattice mismatches in heterostructures. The algorithm's tolerance between experimental and theoretical differences in interplanar distances and angles between them is set to 5%, providing the optimal balance between accuracy in phase identification and robustness to deviations from drift or calibration shifts. Modules 1-4 (Fig. 1) present the sequential process till the indexation of the crystallographic phase and its 3D information. Figure 2 proves this versatility: it shows three examples of paradigmatic nanomaterials and heterostructure configurations in which the workflow univocally identifies the crystal phases and their 3D arrangement. The workflow is capable of detecting highly mismatched interfaces such as cubic defective InSb grown on InP and detecting their relative orientation (Fig. 2.a). [69, 70] It can also identify the closest pure unit cell in a binary compound, like the $Si_{0.3}Ge_{0.7}$/Ge/$Si_{0.3}Ge_{0.7}$ quantum well in Figure 2,b to set the ground of the stoichiometrical refinement that will be addressed later in the workflow's pipeline. [71] Furthermore, showcasing the model's versatility and robustness, in a different materials science domain, low-contrast TEM micrographs of CuTe nanoparticles are indexed and Bragg-filtered. This demonstrates that our phase identification can successfully point at the correct *Pm-3n* phase among the up to 12 different checked candidate CuTe crystal phases (Fig. 2.c). [72]

The robustness of the phase identification is visible in the case studies depicted in Figure 2 and more extensively proved for additional heterostructures, devices, crystal phase types and spatial groups, morphologies, and orientations, in the supplementary information, sections 1.5 "Phase identification" and 2 "Proofs of the automated phase identification". The automated phase identification proposed in this work, grounded in a model-experiment comparison, achieves remarkable accuracy and robustness. Furthermore, its added value lies in its integration into the comprehensive analytical workflow that exploits its output to access further structural and functional insights of devices.

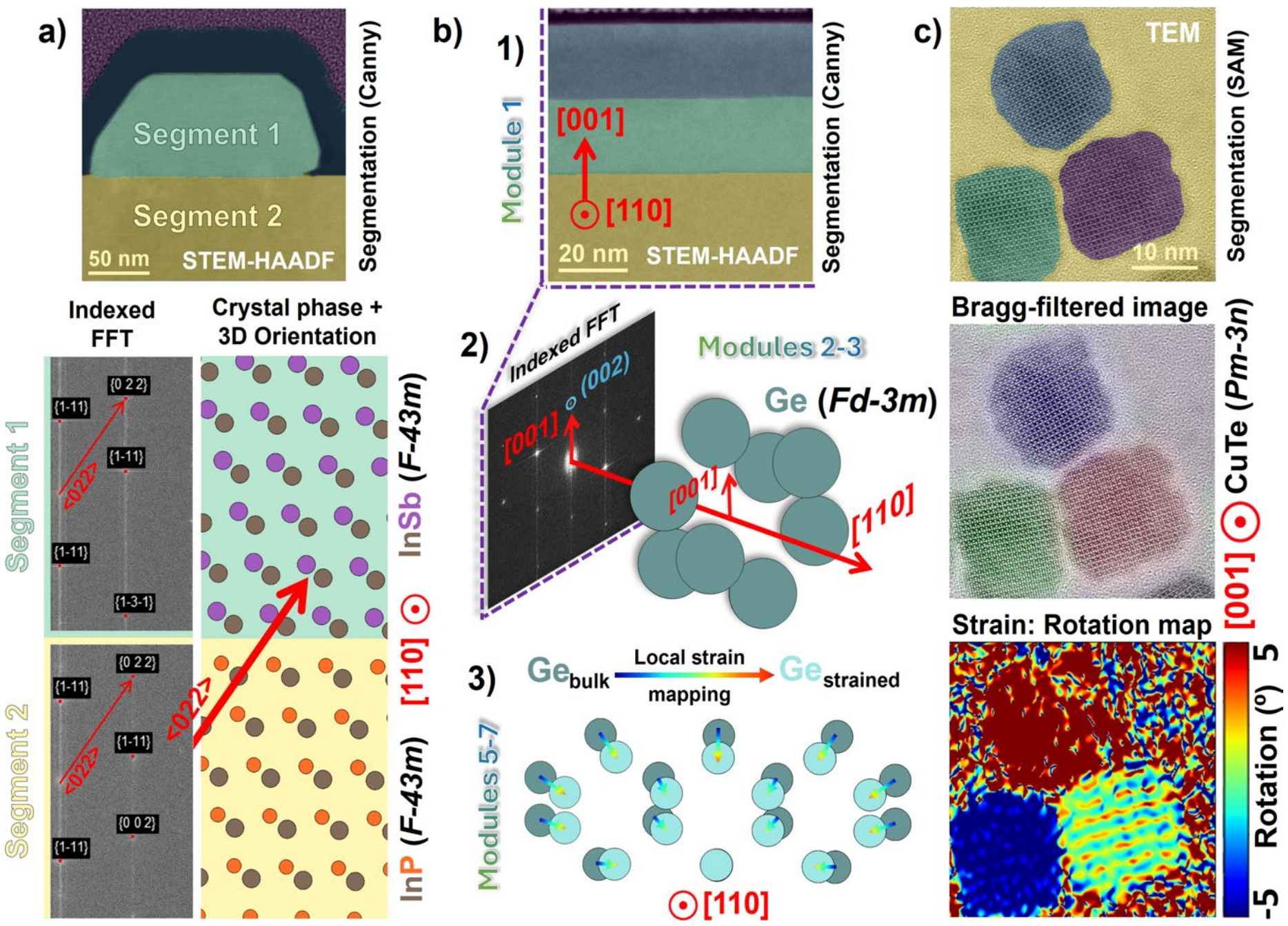

Figure 2: Outputs from the automated workflow for the comprehensive structural characterisation of (hetero)structures and devices. a) Scanning Transmission Electron Microscopy (STEM) image of an InSb nanowire cross-section on an InP substrate oriented along [011] axis. The contour-based segmentation accurately separates the four distinguishable regions of the heterostructure within the lamella (each color is a separated segmented region). The automated phase retrieval indexes the Fast Fourier Transform (FFT) per region and finds the spatial orientation of both constituting phases. [69] b) STEM micrograph of a $Si_{0.3}Ge_{0.7}$/Ge/$Si_{0.3}Ge_{0.7}$ quantum well. [71] 1) The Canny edge segmentation (module 1) and 2) phase identification (modules 2-3) conclude that the three segmented regions identified are formed by Ge with diamond structure, oriented along zone axis [011], as we limited the candidate unit cells to either pure cubic Si or Ge. 3) Modules 5-7 create the 3D atomic model of the Ge bulk lattice (dark green) representing the observed epitaxy, and distort it based on the compressive strain that the Ge in the quantum well perceives due to the bottom $Si_{0.3}Ge_{0.7}$ buffer layer. c) High-resolution parallel beam TEM micrograph of three CuTe cubic particles (top panel). SAM perfectly segments the particles and highlights how the workflow is general and applicable to multiple device morphologies. From the segmentation and the *Pm-3n* phase identification, we can Bragg filter (middle panel) the particles based on their in-(image)-plane relative rotation, as well as computing their local strain, validating their relative rotation (bottom panel). [72]

## 2.4 Low-to-high magnification correlation

The modules described so far operate on single images. Module 4 allows the treatment of image stacks and circumvents the insufficiency of a single image to capture every morphological and structural detail, which is common in large devices of several $\mu m$. The process stacks micrographs of the same device, but with varying experimental parameters like magnification, focus, and sample orientation. These parameters are manually adjusted and defined during the acquisition process. The core idea involves sorting these images by field of view (FOV) and automatically matching them in a chain of template and query images. The mathematical core, multiscale template matching, ensures the adequate pixel size-based scaling of the template-query pairs to maximise their matching based on cross-correlation. The low magnification images are segmented to reveal device morphology taking advantage of the reduced number of pixels per contour. Next, the automated reciprocal space analysis is performed on higher-magnification images containing structural details. This enables mapping crystal phases from atomically-resolved images, but linking them to the lowest magnification images, providing structural information in FOVs where no atomic resolution can be achieved. For instance, we have achieved crystal phase mapping (i.e., identification, 3D orientation, indexation) in FOVs of up to 1197 nm keeping the atomic resolution accuracy. This particular FOV would demand an image of around 20000 pixels per side to be within the Nyquist regime of atomic resolution. [73] We thus replicate the advantages of 4D-STEM acquisitions without venturing into its big data, while also anticipating the future application of this workflow in 4D-STEM. The multiscale matching process is detailed in the supplementary information, section 1.4 “Low-to-high magnification correlation”.

## 2.5 Strain analysis

The combined knowledge retrieved so far (device morphology, local FFT indexation and crystallographic phases) reveals a global and averaged picture of the material atomic arrangement. The workflow automatically maps the existing epitaxial relations, identifying heterojunctions or single crystalline blocks. This is used to automate the Geometrical Phase Analysis (GPA) for relative strain mapping. The method involves selecting a segmented crystalline region as the reference while identifying optimal crystal planes ($g$ vectors) from local crystalline neighbors to compute their joint geometrical phase and the strain. [42, 43, 63, 64] For instance, in the examples showcased in Figure 2, different epitaxial relations are automatically found (see supplementary information, sections 1.6.2 ”Selection of the optimal $g$ vectors pair and mask resolution” and 1.9.3 ”Strain transfer from GPA to atomic model”) . In figure 2.a and b, the workflow gauges a perfect epitaxy with high 10.3 % (Fig. 2.a, and Fig. S23) and no (Fig. 2.b and Fig. S24) mismatch, while in Fig. 2.c, a polycrystalline nature equivalent to the actual rotated single crystals is retrieved (Fig. S30). Further practical details and the resulting automated strain maps from these and other samples can be found in the supplementary information, section 1.6 “Strain analysis: Geometrical Phase Analysis automation”.

The indexed crystallographic planes from the segmented regions are compared to the selected $g$ vectors, and to evaluate local lattice distortions, their surrounding reciprocal space is masked to align with neighbouring indexed crystal planes while balancing spatial resolution and noise. In Figure 2.a, the mask allocates the two neighbouring crystal planes representing the heteroepitaxy (Fig. S23), whereas in Fig. 2.b, it only includes the main plane indicating the homogeneous cell parameter (Fig. S24). In Fig. 2.c the mask opens to allocate three planes, each representing one of the three nanoparticles (Fig. S30), to calculate their relative in-(image)-plane orientation within the same zone axis, which is visible both in the automated Bragg-filtered image and in the rotation map (further details in supplementary information, section 1.6.2 “Selection of the optimal $g$ vectors pair and mask resolution”).

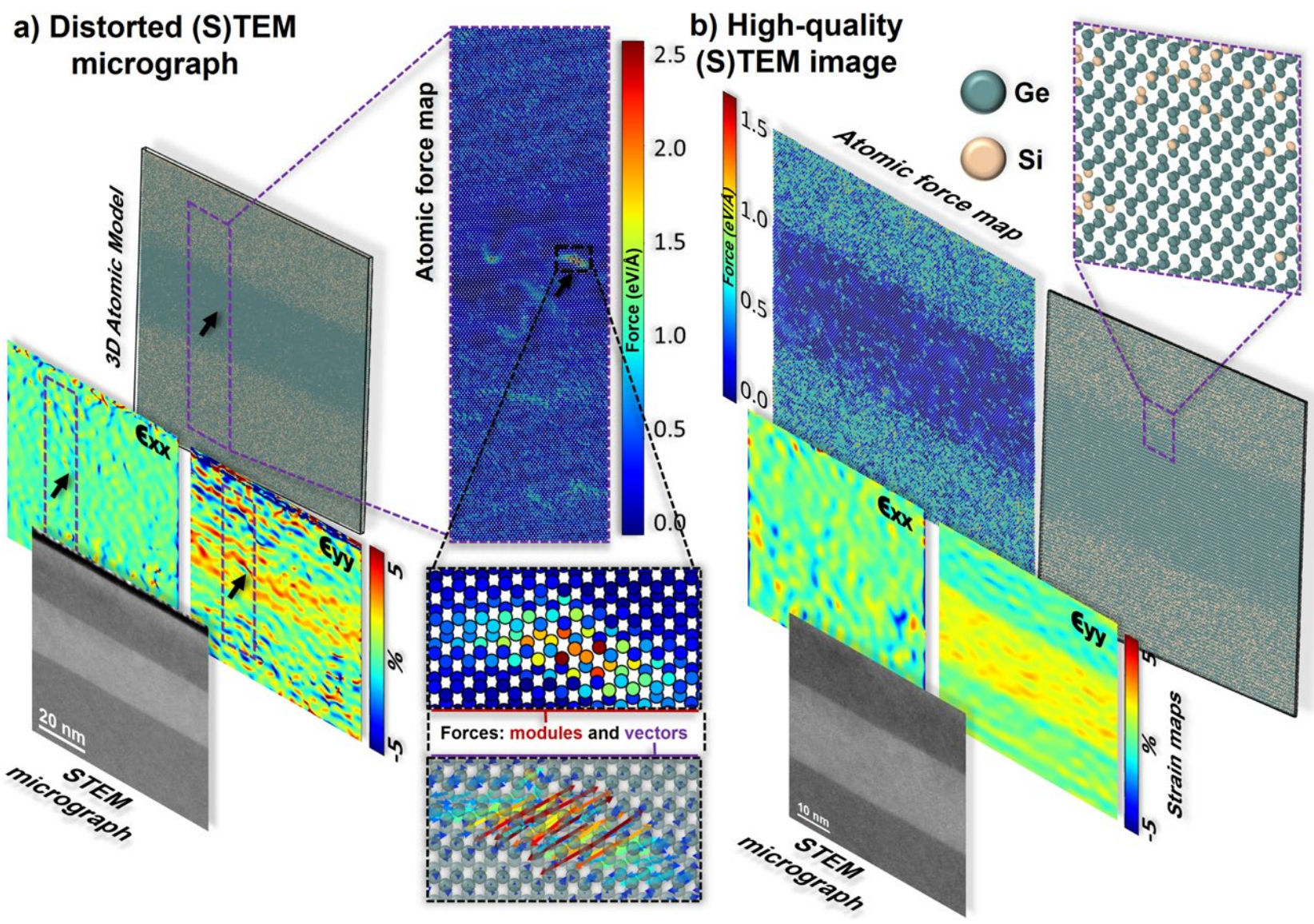

Figure 3: Transfer of atomic-resolution information and relative strain mapping into 3D atomic models (3DAMs). From high-resolution (scanning) transmission electron microscopy ((S)TEM) images, the $\epsilon_{xx}$ and $\epsilon_{yy}$ relative strain maps are computed and its contained information is transferred to 3D atomic models to represent the experimental strain fields in an automated way. The fidelity of the eventual 3DAMs depends on the quality of the experimental (S)TEM micrographs from which they originate: a) Process involving a STEM micrograph of a $Si_{0.3}Ge_{0.7}$/Ge/$Si_{0.3}Ge_{0.7}$ quantum well with scanning artefacts and heterogeneous resolution of the atomic lattice. [71] The effects, turning into uneven strain maps, are transferred into the 3DAM unrealistically distorting the atomic lattice and causing strong atomic force accumulations

(black arrow in the strain and atomic forces maps). b) Process involving a high-quality STEM micrograph of the same device or heterostructure with a smooth and even representation of the atomic lattice and consequently, of the strain. Its transfer to atomic models ends up with a realistic 3DAM where slight heterogeneities in the atomic forces arise due to the placement of the individual Si and Ge atoms in a strained lattice based on local quantitative electron energy loss spectroscopy. The atomic force maps are computed with a Keating model of Si and Ge alloys.

This methodology provides insights into the spatial dependencies of the elastic strain tensor components, the junctions between materials, their structural interaction or epitaxial relations, and the presence of dislocations in their interface (Fig. 4.a, and Fig. S23, S26, S28 and S59). Assessing the relative orientation of crystals and their epitaxial relations is valuable for visually mapping planar defects within the crystal phase, particularly with the automated Bragg filtering in the workflow. In fact, the heteroepitaxial relations automatically found are also used to adjust the Bragg filtering of crystal planes linked to each phase, revealing their spatial distribution. [74, 75] Please refer to the supplementary information, section 1.7 “Bragg filtering automation” for details on the automation of Bragg filtering. Moreover, our automated GPA routine, detailed in section 1.6 “Strain analysis: GPA automation” of supporting information, highlights subtle structural features within the different interfaces of the device as abrupt local variations of the measured strain fields, going beyond the capabilities of the previous crystal phase identification module. This is particularly interesting for the global semiconductor field, including its powerful industry. Overall, it detects defects, dislocations, and alterations from the expected perfect crystal lattice, offering detailed insights that are not easily discernible to the naked eye. For example, the automated rotation map (supporting information, Fig. S23) in the system in Fig. 2.a highlights the stack of dislocations and the presence of a stacking fault. [69, 70] These elements were invisible to the workflow until this step. Furthermore, the quality and smoothness of the GPA maps depend on the quality of the original (S)TEM data, as with the original manual GPA routine. Figure 3 compares the quality of the retrieved strain maps of a Ge quantum well based on image quality and demonstrates how the smoothness of the resulting maps is closely linked to it. This concept is crucial for understanding how this knowledge is transferred to the eventual atomic model, as detailed later (Fig. 3). Similar examples prove the versatility of the method in the supplementary information, section 1.6.5 “Proofs of the automated GPA”.

The retrieved strain maps represent strain relative to the chosen reference region. [42, 43] From them, the translation from relative dilatation to absolute in-plane strain components is immediately computed, as the relaxed cell parameters and their local changes in each segmented region belonging to each phase have been automatically retrieved in the previous steps (supplementary information, section 4.6.2 “Estimation of $1/l_{SO}$”). In fact, it is the combination of these outputs that allows the posterior generation of atomic models that accurately represent the retrieved structural details, as discussed below. However, a deep interpretation of the GPA’s output is a task that still should be done by a trained materials scientist.

## 2.6 Finite Element Model (FEM) building

At this stage, the workflow has collected enough information from the (S)TEM image to, assuming translational invariance along the zonal axis of the image, create 3D models that are representative of the device or material (hetero)structure. At the same time, this assumption constitutes the main restriction of the model building to enable transforming the 2D projection from a micrograph into a 3D model. The model represents the information contained in the imaged atomic columns based on either the identified crystal phases, or spectroscopy acting as a complementary source of local compositional information, which is particularly important in binary, ternary or n-ary compounds such as SiGe alloys.

**FEMs creation:** Our first approach is the automated creation of Finite Element Models (FEM), which describe intricate physical systems using discrete geometric elements. This can get us results that would otherwise be impractical when theoretically considering the entire global system at once. [76, 77, 78] To do so, we create contour vectors from the interfaces that arise between the segmented regions, forming a boundary element model. We encode it into a Graphic Data System (.gds) file,[79] which contains contour information, but also the identified crystal phases and their spatial orientation to provide the correct orientation of anisotropic materials in the continuum FEM. The level of morphological detail of the contour model is easily adjusted by manipulating the number of total nodes dividing the contour. Thus, we trade-off between smoothing curves to reduce aliasing when segmenting downscaled images, and capturing subtler details by keeping more boundary elements.

The workflow-based model creation outperforms traditional approaches to FEM work. It is automated, fast, and does not require manual input for system properties or tedious building of morphology with simple polygons. Moreover, as it is based on experimental data, it is more accurate and representative of the actual heterostructure or device being simulated. However, the introduction of simulation properties (e.g., mesh size and resolution, material properties database, boundary conditions) still requires manual setup in the simulation software of choice. [80, 81, 82] The details on its automation can be found in the supplementary information, section 1.8 “Finite Element Modelling automation”.

**FEM calculation example and discussion:** We subjected the workflow to scrutiny by importing the InSb-InP nanowire cross-section presented in Figure 2 in COMSOL to evaluate its relaxation in physical scenarios of interest represented by different boundary conditions. [69, 70, 80] The results are illustrated in Figure 4.a, demonstrating the straightforward adaptability of simulations based on the proposed workflow. The device’s high mismatch of 10.3% experimentally forces the creation of an array of dislocations in the InSb/InP interface. This effect is observed in the automatically computed strain maps (Fig. 4.a, “Experimental GPA”, and supplementary information, “Proofs of the automated GPA”). As dislocations are an atomic effect, we manually fine-tune the automatic output from the workflow (Fig. S61) to consider them in our con-tinuous simulation model (Fig. 4.a, “FEM strain relaxation”). [83, 84] We also manually define the calculation setup. First, the strain state of the system is incorporated via thermal strain by setting a thermal expansion coefficient of 10% to mimic the mismatch and the elastic component of the relaxation. Second, we represent the experimental array of dislocations as an array of cylindrical elements distributed along the interface, where we impose a fixed displacement of a single epitaxial plane as boundary condition in each (Fig. S61). Adding these *ad-hoc* modifications to the FEM is intuitive and well-integrated in the FEM software. This proves the workflow’s versatility to output a base FEM model that can be manually modified based on the particular needs of the system to represent.

We meshed our geometry with a varying mesh density that is maximised in the dislocations. We applied the boundary conditions sets detailed in the supplementary information, section 4.2 “FEM simulations” to present a common yet fully unresolved issue within materials science and TEM: the effect of lamellae thinning in strain mapping. [85] We compare the unthinned device considering infinite translational invariance with a thinned TEM lamella of 40 nm. This respectively involves applying or not applying a boundary condition of null displacement in the transversal facets of the device. This approach allows us to study the expected difference in the strain relaxation between our measured thinned lamellae and the real device in the wafer. The complete description of the problem is detailed in the supplementary information (section 4.2 “FEM simulations”), but for simplicity, we present the representative component, $\epsilon_x$, in Figure 4.a to compare both scenarios. The first and most obvious observation is the overall larger value of $\epsilon_x$ in the unthinned system. Specifically, when considering the infinitely thick “lamella”, we obtained average values of 11-12 % dilatation in the nanowire. However, in the thinned version, the dilatation only reaches around 9 % in both $\epsilon_x$ and $\epsilon_y$ components (Fig. S63). This implies that the effect of the lamella thinning reduces these components by approximately 22 % $(1 - \frac{9\%}{11.5\%})$. In other words, thinning a 40 nm lamella in the present system releases 22 % of the expected transversal strain through longitudinal relaxation as observed by the lamellae expansion (figure 4.a and supplementary information, section 4.2 “FEM simulations”). Consequently, our experimental TEM measurements would reflect 22 % less strain than the unmodified value we would ideally measure.

It is also interesting to observe from the 3D views of the $\epsilon_x$ and $\epsilon_y$ components (Fig. S63), how the shape of the NW, especially at the kinks on both edges of the interface, unevenly modulates the strain. This highlights the importance of having the actual experimental morphology of the sample rather than just a schematic simplification of the ideal system. However, the comparison with the experimental strain measured with GPA is still discordant with the simulations, which release more strain than the limit of the lattice mismatch. This discrepancy arises because the simulations only consider a stack of dislocations along the transversal dimension and not longitudinally. The $x$ and $y$ components compensate for the remaining unrelaxed longitudinal strain. Notably, the $y$-component (Fig. 4.a) weighs a larger percentage of this relaxation, given that the epitaxial conditions tightly constrain the $x$-component. In summary, with the present workflow, we can intuitively tune our experimental FEM simulation to improve the match of our experimental system. As a consequence, a sample-wise analysis of the studied effect is unlocked, as well as the exploration of the implications of varying lamella thicknesses, among endless possibilities. Further details and discussion about the automated FEM process are available in the supplementary information section 4.2 “FEM simulations”.

## 2.7 3D Atomic Model (3DAM) building

We can go beyond FEMs with 3DAMs and provide a more precise description of the actual device and its structural features. We keep the same translational invariance assumption implying perfect periodicity along the out-of-plane axis, which allows us to build models of any desired thickness. [86, 87] While the atomic description allows for higher theoretical levels in simulations, its building and the actual calculations come at a greater cost in computing time and resources than the FEM. [63, 64, 88, 89, 90, 91] The fundamental concept here is to generate a representative atomic model: a file containing the three spatial coordinates of all atoms constituting the entire device, material (hetero)structure, or a specific region of interest. To achieve this, we require the combination of the output from every previous module of the workflow.

The gathered knowledge from the workflow is first used to populate the volume occupied by the device (defined by the segmentation) with atoms based on the found conventional unit cells and their symmetry operations. [92] These atomic positions are subsequently linked with GPA analysis, particularly with displacement maps. The central concept is the correlation of the displacement maps with the atomic positions to calculate their local atom-wise displacement to the position where they are experimentally found. The key is to build the initial atomic model based on what we refer to as “virtual unit cell”, which is the unit cell whose cell parameters match those of the reference area of our automated strain analysis (Fig. 2b.2). In other words, since displacements are relative to our reference, we need to build everything based on this reference to later apply the modifications (i.e., displacements) relative to it (Fig. 2b.3, dark green Ge atoms displaced to light blue Ge atoms to account for compressive strain). We use the two crystal planes ($g$ vectors) used for GPA to obtain the experimental plane spacing, which we refine at sub-pixel level from the FFT, to compute the resulting virtual unit cell (details in section 1.9.2 “Virtual unit cell calculation” of supplementary information). Note that since only $x$, $y$ displacements fields are extracted from GPA, the estimated out-of-plane interplanar distances of the reference cell are used globally per segment, e.g. in the $Si_{0.3}Ge_{0.7}$/Ge/$Si_{0.3}Ge_{0.7}$ quantum well of Fig 3, the interplanar distances along the zone axis [110] are fixed, according to the value extracted from the reference virtual cell.

The automated strain analysis module is aware of epitaxial relations. Consequently, the construction of the 3DAM will be guided by this knowledge, too. Specifically, epitaxy detection groups two or more epitaxed regions into a replicated single virtual unit cell, from which the atoms are then displaced. For instance, in the $Si_{0.3}Ge_{0.7}$/Ge/$Si_{0.3}Ge_{0.7}$ quantum well displayed in Figure 3.b, a single virtual pure Ge building block is used for the three regions of the device (Fig. 2b.2): the quantum well and the upper and lower buffer layers (details in supplementary information, section 1.9.4 “Compositional information: Spec-troscopic mapping and quantification). [71] Nevertheless, in this case, the resulting atomic model built of only Ge atoms does not consider the local binary stoichiometry of the SiGe alloy yet, as this cannot be inferred from reciprocal space-based phase identification. Instead, local compositional information is obtained from electron energy loss spectroscopy (EELS) to spatially gauge the Si:Ge ratios. As we work with the translational invariance assumption for 3DAM building, this knowledge, which comes averaged throughout the atomic columns, is translated into atomic columns with fractional occupancies (i.e., each atom simultaneously is x% Si and (100-x)%) coinciding with the experimentally found local stoichiometry. In the simplest approximation, these atomic occupancies can then be randomly collapsed into definite occupancies based on the homogeneity of the quantitave EELS maps (i.e., either Si or Ge) (Fig. S38). The obtained 3DAM can be seen in Figure 3. Avenues to improve some of the approximations made above (constant out-of-plane cell parameter and random occupations) are discussed in section 2.8. [71] The process to correlate and integrate EELS with the workflow is based on the multiscale template matching of the quantitative EELS maps with the main analysed image (Fig. S38), similar to section 2.4, and is detailed in the supplementary information, section 1.9.4 "Compositional information: Spectroscopic mapping and quantification".

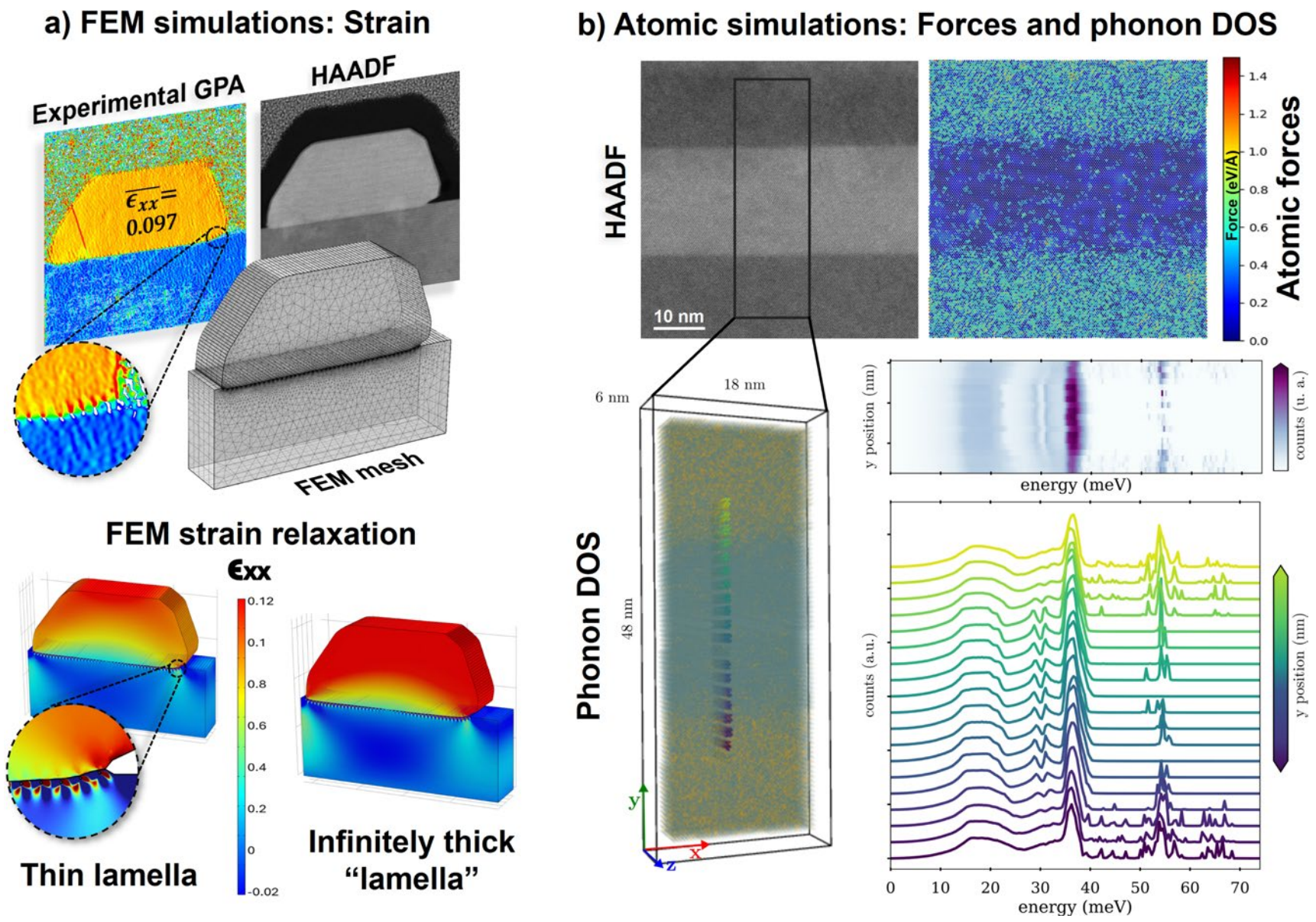

Figure 4: Multiscale computational materials science: Finite Element Model (FEM) and atomic simulations from the workflow's outputs. a) FEM simulations of an InSb nanowire grown on InP. [69] The segmentation extracts its morphology to make a 3D geometrical FEM mesh of the experimental structure. As an example application, it unlocks computing strain relaxations and gauging the effect of lamellae preparation even in the presence of fine anomalies like dislocations (zoom-

in) and subtle morphological features. b) Atomic simulations from the 3D Atomic Model (3DAM) of a $Si_{0.3}Ge_{0.7}$/Ge/$Si_{0.3}Ge_{0.7}$ quantum well. [71] The Keating model is used to calculate the atomic forces, which are greater in the upper and lower buffer layers due to the presence of Si (in accordance with the expected occupancy) in the pure germanium virtual lattice. Form the force constant matrix, the phonon density of states is calculated. The phonon-density of states (DOS) map along the y-axis of the heterostructure shows the modulation of the optical mode of Si at 55 meV when going through the variations of the local Si content.

The situation slightly differs if the epitaxy arises from highly mismatched device regions, separated by the segmentation model at the beginning of the workflow. In these cases, the crystal planes of both regions appear as separated and distinguishable pairs in the FFT, which enables their separate crystal phase identification (see details in supplementary information, section 1.9.5 "Compositional information: Segmentation and symmetry equivalences"). Multiple strain calculations are sequentially computed by varying the reference position in every material participating in the epitaxy. Every virtual atomic block is displaced independently based on its own virtual reference unit cell, and then the segmented regions are fitted and merged as puzzle pieces to form a 3DAM representing the full device (details in supplementary information, section 1.9 "Atomic model building", specifically 1.9.3 "Strain transfer from GPA to atomic model"). This pertains to the InSb-InP structure displayed in Figures 2.a and 4.a. Its 3DAM formation process, and discussion on how dislocations, stacking faults and local strain modulations are captured can be found in the supplementary information, section 3 "Proofs of the full workflow", and in Figures S23, S31, S37, S43 and S59. [69, 70]

Achieving a high-quality representative 3DAM is highly sensitive to the micrograph acquisition process. The quality of the reconstruction is closely linked to the micrograph quality and its subsequent assessment of the automated strain maps retrieved, as showcased in Figure 3. Thus, as in classic GPA, scanning artefacts, noise and undersampling can affect the smoothness of the resulting strain maps. Cumulatively, these issues manifest as model artefacts in the reconstructed 3DAMs (Fig. 3.a). Therefore, although phase identification remains robust even with flawed data, accurately describing the positions of the atomic columns requires processing from high-quality data. (Fig. 3.b). Intuitively, providing atomically precise outputs requires a likewise atomically precise flawless starting point. A first way to validate the quality of inputs for 3DAM building is through interpeting the strain maps, as they are highly sensitive to imaging artefacts.

Figure 3 demonstrates how uneven strain maps ($\epsilon_{xx}$, $\epsilon_{yy}$) with significant local modulations (indicated by the black arrow) from a distorted STEM micrograph can yield defective 3DAMs. This can be independently and sensitively detected by an atomistic computation of forces (see calculation details in the next section): as seen in Figure 3.a, the model constructed from low-quality data displays a region of non-physically high forces (indicating out-of-equilibrium atomic positions). Conversely, high-quality data, represented by smooth GPA maps, leads to representative 3DAMs, whose atomic force maps do not display such non-physically high forces (Fig. 3.b). This illustrates that calculating atomic forces serves as a complementary verification of the 3DAM reconstruction quality. Non-physical local peaks in the atom-wise forces can indicate imaging artefacts. Double-checking them with the local image quality and the strain maps themselves validates them. In the next section, we discuss additional approaches to ensure quantitative data quality for artefact-free 3DAM construction.

Equivalently, these effects caused by imaging artefacts such as the one pointed by the black arrow in Figure 3 could also be seen as the result of applying a custom displacement field to the actual device to, *in silico*, test its impact on the properties of a potential future device candidate (i.e., reverse engineering). Such approaches, using deliberate structure modifications, are grounded in the experimental simulations enabled by the proposed workflow and the concept of creating a reliable digital twin of the device under study.

Overall, these cases demonstrate the workflow's capabilities in generating 3DAMs that would be impractical to construct manually due to the complexities of local atom-wise displacements. Section 3 "Proofs of the full workflow" in the supplementary information contains further examples and the extent of currently supported materials science scenarios.

In addition to providing a detailed structural understanding of the heterostructures functionality, these 3DAMs open the door to improved understanding of material and device physics, as experimentally determined inputs for atomistic simulation methods.

## 2.8 Simulation of functional properties

The final module of the workflow aims to compute the properties determined by the specific atomic arrangement of the device by considering different level of theory description of the 3DAM. We can simulate functional properties with a precision that is only achievable when considering the discrete atomic nature. We can validate the structures and perform quantitative (S)TEM analyses through immediate linear or multislice (S)TEM simulations of the resulting 3DAM (Fig. S60). For instance, we present the results of a linear STEM image simulation of a Ge quantum well with varying lamella thicknesses. This demonstrates the ease of performing quantitative checks such as focal series or varying depth of focus STEM analyses, as well as more detailed evaluations of the effects of zero-point vibrations in micrographs (supplementary information, section 4.1 "STEM simulation details").

The 3DAMs can also serve as inputs for simulation methods such as molecular dynamics or *ab initio* calculations, enabling the computation of functional properties like electrical and thermal conductivity, electronic and phononic density of states (DOS), band gaps, dispersion relations, atomic forces, and other properties essential for understanding complex devices like quantum wells, hybrid nanowires embedded in quantum networks, and catalytic particles. This is precisely what we have done with the particularly interesting $Si_{0.3}Ge_{0.7}/Ge/Si_{0.3}Ge_{0.7}$ quantum well discussed throughout the article (Fig. 4.b and Fig. 5). [71] Its remarkable properties as a platform for hosting singlet-triplet spin qubits make it an ideal candidate for evaluating and envisioning the potential of its representative 3DAMs as digital twins, allowing us to explore its physics in greater depth.

We first calculated the forces acting on each atom in our 3DAMs. This can be accomplished using a number of different methods, ranging from first-principles calculations such as Density Functional Theory (DFT) to empirical classical force fields. [93, 94, 95] DFT does not contain empirical parameters, and therefore does not require fitting to previous experimental or theoretical data. It boasts high predictive power and provides very accurate force calculations. However, its computational cost is significant, specially for systems with a very large number of atoms, such as the 3DAMs considered here, making it unpractical for the purposes of this work. Nevertheless, reduced scaling DFT algorithms [96, 97] and access to massively parallel computing resources have made these calculations more feasible, and we plan to incorporate them in future evolutions of our workflows. On the other hand, empirical classical force fields have relatively simple functional forms based on atomic positions, and are fitted to known data, but they only perform well for systems similar to those used for fitting. The simple functional form makes them computationally inexpensive and capable of being applied to large atomic system. In this work, we utilise a simple Keating model,[44] specifically developed for mixed SiGe systems [98] which strikes a good balance between accuracy and computational efficiency. Using the Keating force field model, we can compute atomic forces for systems of the order of 10**5** atoms in just seconds on a desktop computer.

Fig. 3.a illustrates the atomic forces computed using this Keating model, highlighting its ability to identify artefacts in the construction of the 3DAM from experimental STEM data. The unexpectedly large forces in the region marked by the black arrow stem from a scanning artifact that is difficult to detect in the 3DAM visually. By mapping this position back to the strain map, we can observe local imperfections in the atomic columns, such as a slightly off-axis sample orientation and a damaged lamella. Thus, these force calculations provide a quick and cost-effective method to screen the models and to assess their quality: large forces typically indicate artefacts and faulty models.

In Figures 3.b and 5.b, we present the forces obtained for a 3DAM derived from high quality STEM data, where the forces on all the atoms are small, validating the model. Interesting information can be extracted from these results. Notably, very small forces are observed in the Ge-rich region of the quantum well, whereas larger forces are found in the upper and lower Si-containing regions. This is due to the assumptions done in the construction of the 3DAM, where pure Ge is taken as a reference. When the 3DAM is built placing actual Si atoms in the Ge lattice to match the experimental position-dependent concentration, pairs of different species as first neighbors are expected to have different interatomic distances from the nominal Ge-Ge one, but this is not reflected in the model, thus producing forces in those regions where Si-Si or Ge-Si pairs occur (supplementary information, section 4.3 “Atomic forces and relaxations - Keating model”). This seldom happens in the Ge-rich quantum well, as there are very few Si atoms, but is much more frequent in the buffer layers containing Si, therefore increasing the average atomic forces in those regions.

The computed atomic forces could also be utilised to further refine the structural 3DAM model, yielding more realistic interatomic distances in the regions that deviate from pure Ge stoichiometry. This refinement can be easily achieved by adjusting the positions of the atoms according to the forces until the total energy is minimized, either through a minimization algorithm such as conjugate gradients, or via molecular dynamics with force quenching. [99, 100] We have applied this approach for the model shown in Figure 5.b, resulting in a structure closely resembling the original 3DAM, but with more accurate interatomic distances for Si-Si, Si-Ge and Ge-Ge pairs (see Supplementary Information, section 4.3 “Atomic forces and relaxations - Keating model”). For simplicity, we have not used this refined model further in the calculations of other physical properties shown below, although more realistic structures would yield more accurate results. The same methods may in fact be used to refine the atomic positions in the out-of-plane direction, but this is likewise out of scope of the present work.

The Keating model can be further used to extract additional information about the system. We can calculate the force constant matrix, which consists of the second derivatives of the energy with respect to the atomic positions. From it, we can derive the dynamical matrix and the harmonic vibrational frequencies and modes. While computing the force constants is only marginally more computationally intensive than calculating the forces, and can also be performed in very little time in desktop computers. Computing the vibrational modes from the force constant matrix, though, requires significantly more computational effort, as it involves the diagonalisation of the dynamical matrix, which becomes quite large for systems with many atoms, such as those considered here. To address this challenge, we have developed an alternative algorithm based on Green’s functions (described in supplementary information, section 4.4 ”Phonon DOS - Keating model”), that allows us to compute the vibrational Local Density of States (vLDOS) in selected regions of our 3DAMs with very little computational effort.

Figure 4.b displays the vLDOS for the high-quality 3DAM discussed earlier, across different regions of the device. Specifically, we calculate the vLDOS for columns of atoms at various locations along the variant $y$ axis (see supplementary information, section 4.4 “Phonon DOS - Keating model”). The vLDOS map reveals how local chemistry influences the acoustic and optical bands of the alloy. The vLDOS map of the quantum well is close to the Ge bulk modes, with main optical peak around 37 meV and minimal signal of higher frequency modes from the presence of a few Si atoms in that region. In contrast, significant differences are observed in the acoustic and optical modes within the more Si-rich buffer zones, where distinct Si-Ge and Si-Si stretching optical bands appear at higher frequencies (around 55 and 65 meV, respectively). This information is invaluable for assessing relevant physical properties of the studied device, such as conductivity and transport characteristics of quantum devices, as well as the figure of merit for thermoelectrics. Additionally, the spatial mapping of the vL-DOS would facilitate direct comparisons with experimental STEM vibrational spectroscopy. [101, 102]

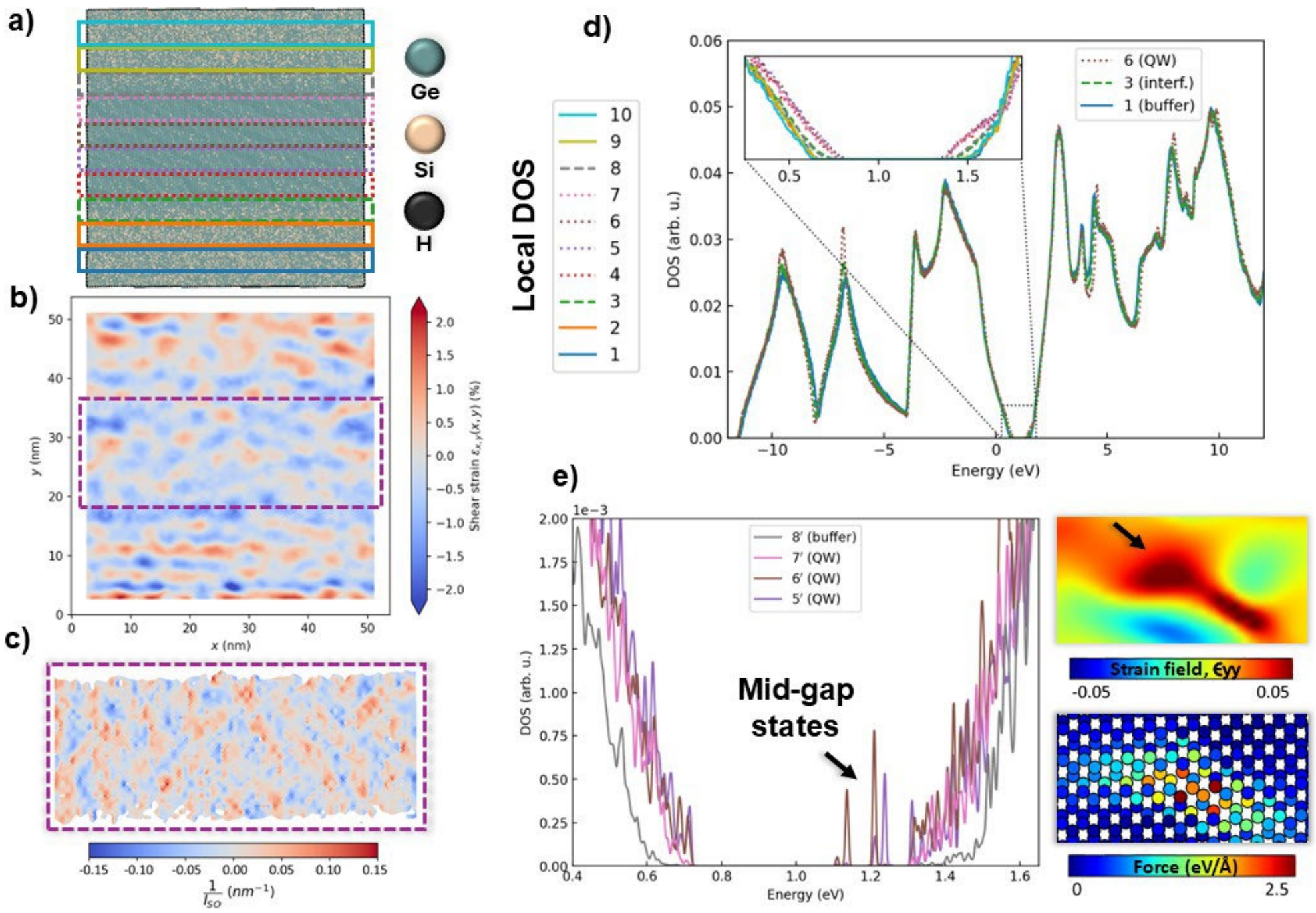

Figure 5: Local electronic density of states (DOS) of a $Si_{0.3}Ge_{0.7}$/Ge/$Si_{0.3}Ge_{0.7}$ through tight-binding atomic simulations. [71] a) Built upon the proposed workflow, the 3D atomic model (3DAM) used for the simulation is divided into boxes where the local DOS is computed. The 3DAM is passivated with H atoms to complete atomic bonding at its edges. b) The automation to compute the experimental shear strain permits us to map (c) crucial local properties for the spin qubits such as the inverse spin-orbit length at high spatial resolutions. d) Local electronic DOS profiling of the quantum well as depicted in panel a). The zoom-in showcases a clean band gap, displaying a valence band edge which is higher within the well and lower in the buffer layers, as expected. e) Zoomed-in near band gap electronic DOS for an atomic distortion centered in the defective model of the quantum well (Fig. 3.a, Fig. S68, S69), in the equivalent regions displayed in a) and d). The distortion can be interpreted as the application of a custom strain accumulation indicated by the black arrow to the 3DAM digital twin of the Ge quantum well device (Fig. 3). The atomic distortion altering the regular lattice in the pm range raises mid-gap states potentially harmful to the transport and quantum properties of the device.

We now turn our attention to exploring the electronic properties of our 3DAMs. For this purpose, we can use methods ranging from first-principles DFT to empirical, simplified electronic Hamiltonians. Similar to our approach for vibrational properties, we opt for simplified and relatively inexpensive models rather than the more accurate but computationally expensive DFT methods. In particular, we adopt an established $sp^3d^5s^*$ tight-binding model for SiGe alloys, which incorporates the effects of strain and spin-orbit coupling. [103] This tight-binding model takes the atomic positions as specified by the 3DAM as input and yields a Hamiltonian that describes the electronic properties of the system. Once this Hamiltonian is established, various atomistic theoretical methods can be applied to extract sample properties, including sparse diagonalisation, non-equilibrium Green's functions (NEGF) for device simulation, or kernel polynomial methods (KPM). [45] The computational cost of KPM scales linearly with the number of atoms, making it feasible for very large sample sizes.

We demonstrate this workflow capability through a DOS calculation using KPM and an investigation of spin-orbit coupling features related to the deformation fields experimentally obtained from our workflow. For DOS calculations, we consider two different model cells derived from distinct 3DAMs. The first supercell, obtained from the purple dotted square in Fig. 3.a (which presents imaging artefacts), consists of approximately 800.000 atoms. The second supercell, extracted from the high-quality data of Fig. 3.b, contains about four million atoms (approximately 80 million spin-orbitals). In both cases, we compute averaged local densities of states (LDOS) in spatial regions profiling the 3DAM along the growth direction, to explore the corresponding changes in electronic properties. For the large cell, these regions are shown in Fig. 5.a. For details on the smaller cell regions, please refer to the supplementary information, section 4.5.2 "Results of the experimental/tight-binding/Keating correlation", along with figures S67 and S68.

In the smallest model cell (Fig. 3.a, dotted purple rectangle), we observe sharp DOS peaks within the electronic gap (Fig. 5.e), which originate from regions inside the quantum well. These peaks are linked to the imaging artefacts discussed earlier. By employing sparse diagonalisation techniques, we can extract electronic wavefunctions at specific energies. We do so for the in-gap peak at approximately 1.13 eV to find that the resulting wavefunction to be strongly localized on the identified imaging artefact, consistently with atomic force calculations (Fig. 3.a, Fig. 5.e, and supporting information, section 4.5.2 "Results of the experimental/tight-binding/Keating correlation"). This suggests that if a Ge quantum well experiences the depicted local strain peak or distortion (indicated by the black arrow), it may exhibit such localised mid-gap states.

The largest model cell is presented in Figure 5.a, illustrating the disordered distribution of Ge and Si atoms across the qubit device. We plot the LDOS over the entire spectral range of the system, along with a zoom-in near the band gap (Fig. 5.d and inset), highlighting the variability of the DOS across ten regions profiling the 3DAM vertically. Of particular interest to hole spin qubit physics is the behaviour of the valence band edge, observable in the inset. We can distinguish three types of regions: buffer regions (1, 2, 8, 10 – solid lines, $Si_{0.3}Ge_{0.7}$), interface regions (3, 8 – dashed lines) and quantum well regions (4, 5, 6, 7 – dotted lines, $Si_{0.03}Ge_{0.97}$), both in the atomic model and in their corresponding DOS signals. [71] We observe that the valence band edge is found at a higher energy within the quantum well compared to the buffer region, while intermediate energies are likely to be found along the interface. This observation aligns with the expected band alignment, as the quantum well is designed to provide a potential well for holes in the Ge portion. Although the conduction band edge is not directly relevant to the operation of this hole spin qubit, it is discussed in the supplementary material, section 4.5 "Electronic DOS - Tight-binding simulations".

Additionally, it is important to highlight the compatibility of our workflow with continuum methods, such as k·p methodologies, which serve as powerful tools for investigating nanostructures. While these methods may lack atomic accuracy, they are well-developed and can offer advantages in terms of reduced computation times and improved interpretability of results. Recent studies have indicated that inhomogeneities in strain and composition across the interfaces in planar SiGe quantum wells could influence the active spin-orbit mechanisms in the system, thereby affecting qubit performance. Better understanding and quantitative estimation of these inhomogeneities are crucial for optimising qubit systems. [6, 104].

Figure 5.b presents the experimental map of absolute shear strain across the sample, revealing significant inhomogeneities. Abadillo-Uriel and co-authors demonstrated that such inhomogeneities promote linear-in momentum spin-orbit interactions. They provide explicit corrections to the minimal k·p Luttinger-Kohn Hamiltonian for all strain-induced spin-orbit interactions, which depend on the Si and Ge Luttinger parameters, deformation potentials (which are tabulated), and derivatives of the strains. These can be evaluated for experimental structures using the proposed workflow.[104] To illustrate this, Figure 5.c displays the inverse spin-orbit length, which serves as a metric to exemplify the strength of these effects in the quantum well region of the heterostructure. For further details on the simulations conducted and their conclusions, please refer to the supplementary information, section 4, "Experimental simulation of key functional properties".

## 2.9 Full Workflow wrap-up

The previous sections traced the workflow's journey from single electron micrographs to simulated properties. We utilised a SiGe-based quantum device known for its exceptional performance in spin qubit computing as an example to validate and demonstrate our findings, particularly due to its promising potential for future improvements and device integration. [71, 105] The intermediate results of the workflow, as schematised in Figure 1, are presented for this device in Figure 2.b (modules 1-7), Figure 3 (modules 5-7), and Figures 4.b and 5 (modules 7-8). Together, these figures provide a comprehensive view of the expected local and global outputs of the workflow. Furthermore, we have validated the workflow with various device types and material configurations representing a broad spectrum of scenarios within materials science, including different SiGe quantum planar devices, hybrid nanowires integrated in quantum networks, and even nanoparticles for photonics and photovoltaics. [69, 72, 106, 107, 108] A detailed description of these validations can be found in the supplementary information, section 3 “Proofs of the full workflow”.

A central focus of our research has been to make the workflow as general as possible, resulting in what we believe to be the most comprehensive data analysis workflow available in the field. However, there is significant room for improvement. For instance, the resulting 3DAMs from high-quality micrographs are robust and representative when elastic strain mechanisms are present. However, first, they are limited by the translational invariance assumption, and second, they currently demand a manual adjustment in capturing strong plastic deformations such as stacking faults, which are only partially represented. Thus, improvements can occur both in enhancing individual modules and in expanding the workflow by integrating additional analytical modules. This expansion will create a larger, evolving framework that can progressively address more materials science scenarios, ensuring its ongoing relevance.

The workflow’s potential and its ability to unveil physical knowledge in a theoretical exploratory manner extend beyond its mere improvements. It facilitates the optimisation of devices with promising prospects, such as the SiGe heterostructures for spin qubits described herein. The attained experimental fidelity in 3D devices and heterostructures unlocked accessing materials and functional properties influenced by the experimental nuances of the actual devices. For instance, the distortion depicted in Figures 3.a and 5.e can be interpreted as applying a custom strain field to the device’s digital twin to assess its *in silico* DOS changes, whose correlation with experimental functional measurements remains as a future perspective. With the resulting knowledge, these experimental simulations are thought to guide the specifications of the next generation of devices. In fact, the results of these simulations could directly come from AI models that directly output physical properties of interest given the original input image, although this is beyond the scope of the present manuscript.

This advancement enhances the predicting power of digital twins, enabling hypothesis testing for custom strain fields, compositional profiles, or isotopic configurations for new device candidates. These steps toward risk reduction in the materials research cycle pave the way for reverse engineering and cost-effective device optimisation, even when dealing with subtle phenomena such as quantum effects.

# 3 Conclusions

We have introduced a workflow that autonomously translates electron micrographs into 3D models suitable for theoretical analysis. This approach provides a rapid, accurate, and comprehensive structural description of imaged heterostructures and devices. Furthermore, it utilises this structural information to generate realistic models, either finite element or atomic, that empower theoretical simulations aimed at extracting functional properties (such as stress fields, strain relaxation, forces, and phonon and electronic DOS) in the final device configuration (digital twin). Our work demonstrates reproducibility and validity across multiple examples and scenarios showcased throughout the main text and supplementary information. For instance, complete (S)TEM data analysis which could take days of expert time, can now be automatically performed in a matter of minutes. However, constructing the 3DAMs through local displacements is currently more time-consuming, taking a few hours, although it yields invaluable information.

Crucially, we believe this workflow establishes a new paradigm in automated data analysis for experimental techniques. The underlying concept can be extended to various methods, with scanning probe and atomic force microscopies being prominent examples, especially for 2D materials. [9, 41, 109, 110, 111, 112, 113] We envision its potential widespread adoption within the microscopy community, both in academia and in industry. For instance, the semiconductor industry would greatly benefit from this robust workflow and its independent and flexible modules. The tool's automated, human-bias-free and user-friendly nature is essential for overcoming the long-standing limitation of TEM: the difficulty in achieving statistical significance. This involves measuring the same properties across multiple devices or samples to minimise its statistical uncertainty, which requires the high-throughput unlocked by the workflow. Moreover, combining this workflow with ongoing developments in (S)TEM data analysis and automation for data acquisition is pivotal in transforming TEM into a high-throughput analytical technique, thus accelerating scientific discoveries. [46, 114]

There is substantial room for improvement, refinement,integration, and inclusion of AI in the modules that currently do not make use of it. Pre-cleaning steps such as denoising, drift correction, and scanning artefacts correction could be easily added to enhance robustness. [36, 115, 116, 117] Additionally, there is ample opportunity to expand into other acquisition modes, like 4D-STEM, or to incorporate spectroscopy-related functionalities in both low and high-loss regimes. [118, 119, 120, 121, 122] Automating the final modules that currently involve manual processing is also feasible, specially through physics-aware AI trainings. [90, 91, 123, 124, 125] For instance, by grouping condition-specific simulation profiles to expedite the process. While these considerations extend beyond the present scope, they represent exciting avenues for future research.

Finally, we want to address the human-AI interaction aspect. We envision an end-to-end general generative model capable of automatically retrieving everything from experimental devices: from acquisition to data analysis. We have laid a cornerstone for this vision, and while it may be some time before such a model is realised, we believe that humans should not fully abandon the intermediate steps of the workflow. The advantages of automation should be complemented by human oversight. Certain steps, such as post-processing segmentation or peak identification, could benefit from validation by trained individuals, enriching the analysis. Additionally, establishing control points or security checks from the outset is beneficial and does not hinder the original aim of revolutionising the electron microscopy analysis as we know it to date.

# 4 Acknowledgements

ICN2 acknowledges funding from Generalitat de Catalunya 2021SGR00457. The authors thank support from the project AMaDE (PID2023-149158OB-C43), funded by MCIN/ AEI/10.13039/501100011033/. This study is part of the Advanced Materials programme and was supported by MCIN with funding from European Union NextGenerationEU (PRTR-C17.I1) and by Generalitat de Catalunya. We acknowledge support from CSIC Interdisciplinary Thematic Platform (PTI+) on Quantum Technologies (PTI-QTEP+). This research work has been funded by the European Commission – NextGenerationEU (Regulation EU 2020/2094), through CSIC's Quantum Technologies Platform (QTEP). ICN2 is supported by the Severo Ochoa program from Spanish MCIN / AEI (Grant No.: CEX2021-001214-S) and is funded by the CERCA Programme / Generalitat de Catalunya. Part of the present work has been performed in the framework of Universitat Autònoma de Barcelona Materials Science PhD program. IPH acknowledges funding from AGAUR-FI scholarship (2023FI-00268) Joan Oró of the Secretariat of Universities of the Generalitat of Catalonia and the European SocialPlus Fund. M.B. acknowledges support from SUR Generalitat de Catalunya and the EU Social Fund; project ref. 2020 FI 00103. This study was supported by EU HORIZON INFRA TECH 2022 project IMPRESS (Ref.: 101094299). Authors acknowledge the use of instrumentation as well as the technical advice provided by the Joint Electron Microscopy Center at ALBA (JEMCA). ICN2 acknowledges funding from Grant IU16-014206 (METCAM-FIB) funded by the European Union through the European Regional Development Fund (ERDF), with the support of the Ministry of Research and Universities, Generalitat de Catalunya. ICN2 is a founding member of e-DREAM.[126] SR is also supported by MICIN with European funds NextGenerationEU (PRTRC17.I1) funded by Generalitat de Catalunya and by 2021 SGR 00997. PO acknowledges support from the EU MaX CoE (Grant No. 101093374), Grants No. PCI2022-134972-2 and No. PID2022-139776NB-C62 funded by the Spanish MCIN/AEI/10.13039/501100011033 and by the ERDF, A way of making Europe. and by Generalitat de Catalunya (Grant No. 2021SGR01519).We thank the Catalan Quantum Academy for support. The authors acknowledge Dámaso Torres for his support in designing the graphical material.

# 5 Supplementary Information

Supplementary information is available under request to the corresponding authors.

## 6 Open data

The data and code used in the present work can be found in the following repository (a private GitHub repository available, to open to the public as soon as the manuscript is accepted).

## 7 Conflicts of interest

The authors have no conflicts to declare.

## 8 Authors contributions

M.B. and J.A. conceived the idea and wrote the manuscript. M.B. led the conception of the workflow, all modules and their interconnection. M.B. and I.P-H developed the segmentation and peak finding modules. I.P-H developed the GUI. M.B., E.R. and V.G. developed the phase identification, GPA automation and 3DAM building. T.G. and S.R. led the tight-binding simulations. C.C. and P.O. led the Keating model-based simulations. P.H.K. and V.G. led the FEM simulations. M.C.S and E.R. led the STEM simulations. Y-M.N assisted with the tight-binding calculations. M.B.E. and G.M. assisted in the peak finding deep learning models. J.A, S.M.S and M.B. acquired the experimental (S)TEM data. G.I, A. C., G.K., G.S. and P.K. provided the materials platforms under study and the discussion on their functional properties and applications. All authors contributed to the manuscript and overall discussion.